\documentstyle[12pt,aaspp4]{article}

\def\ni{\noindent}

\def\emfb{\overline{\mbox{\boldmath ${\cal E}$}} {}}

\def\beq{\begin{equation}}
\def\ee{\end{equation}}
\def\lsim{\mathrel{\rlap{\lower4pt\hbox{\hskip1pt$\sim$}}
    \raise1pt\hbox{$<$}}}
\def\gsim{\mathrel{\rlap{\lower4pt\hbox{\hskip1pt$\sim$}}
    \raise1pt\hbox{$>$}}}

\def\ts{\times}

\def\bfv{{\bf v}}

\def\bfb{{\bf b}}

\newcommand\oS{{\overline C}}
\newcommand\cc{{c}}
\newcommand\oQ{{\overline Q}}

\begin{document}

\setcounter{equation}{0}

\centerline{\large\bf A New Approach to Turbulent Transport of a  Mean Scalar}
%\centerline{\large\bf on Turbulent Magnetic Dynamo Spectra}

%\bf The Effect of Fractional Kinetic Helicity on Small-Scale Dynamos:}
\medskip
%\centerline{\bf Application to the Galactic Magnetic Field} 

\author{Eric G. Blackman \altaffilmark{1} and George B. Field\altaffilmark{2}}
\affil{1. Department of Physics \& Astronomy and Laboratory for
Laser Energetics, University of Rochester, Rochester NY 14627}
\affil{2. Harvard-Smithsonian Center for Astrophysics, 60 Garden St.
Cambridge MA, 02138}

\centerline {(submitted to Physics of Fluids)} 

\begin{abstract} 
%Triple correlations must be included in any realistic 
%approach to turbulent processes since they provide the nonlinear 
%turbulent cascade. 
We develop a simple mean field approach to the transport 
of a passive scalar for which the fundamental equation 
is a second order differential equation in the transported quantity, not 
a first order equation.
Triple correlations are included, as they must be for any realistic
description of turbulence.  No 
correlation time enters the theory, only an eddy turnover time.
The approach is simpler than standard approaches which
incorporate triple correlations, 
but more realistic than Gaussian or short correlation 
time closures which do not. 
A similar approach has  proven useful in magnetohydrodynamics.

\end{abstract}

\centerline{PACS codes: 
47.27.Qb; 92.10.Lq; 94.10.Lf; 95.30.Qd;96.20.Br;83.50.Xa}

%95.30.Qd;  98.38.Am; 52.55.Ip,
%47.27.Eq; 
%52.30.Cv; 98.35.Eg; 
%96.60.Hv}92.10.Lq
%\bigskip
%\tableofcontents
%\listoffigures

%\section{New stuff}

%\section{Introduction}

%{\it Introduction-}
Transport of a passive scalar
in a turbulent flow is a classic problem in the physics of fluids [e.g. 1-10]. 
Most recent 
formal work has focused on spectral scalings, non-Gaussian statistics,
and high order  
correlations of a passive scalar fluctuations [e.g. \cite{shraiman,falkovich}]
rather than on the dynamical
theory  of the evolution of a mean scalar field 
[1,11-14].
%\cite{elperin2000,elperin2001,avellaneda92}].  
However, understanding how a mean scalar such as average concentration  
evolves in a turbulent flow 
is important for a broad range of problems  such as the 
transport of pollutants and cloud 
droplets in atmospheric turbulence [\cite{pasquill83}],  
dust transport in planet forming disks  [\cite{hodgson98}], 
and particle transport in industrial flows [\cite{baldyaga99}].

A proper mean field theory requires a closure 
%[e.g.\cite{zeldovich88,zeldovich90,mccomb,frisch}] 
which allows the evolution 
of the mean scalar to be represented by a finite set of 
solvable realistic equations.
If a closure can be tested and proven useful for 
a the  problem of mean scalar transport, further application and testing  
in fluid dynamics  and  magnetohydrodynamics (MHD) is  motivated.  
For problems amenable to a mean field treatment,  
a simpler closure than that needed for a full spectral treatment may be 
sufficient. Here we derive a simple practical approach  
to the Eulerian transport of a mean scalar in a turbulent flow 
that differs from standard treatments [e.g. 1-8].
but whose analogy has proven effective in mean-field magnetic 
dynamo theory [\cite{bf02}].
We also discuss the physical motivation for the approach and
the relation to previous closures.

To study mean scalar transport,  
we start with the basic conservation equation [\cite{taylor21,mccomb,lesieur}] 
for the concentration of a passive contaminant, $C$. 
For incompressible flow, we have
\beq
\partial_t C= -\bfv\cdot\nabla\oS -\bfv\cdot\nabla \cc,
\label{1}
\ee
where $C=c +\oS$, and  $c$ and $\oS$ are 
the fluctuating and mean components of $C$ respectively.
The velocity has only a fluctuating component $\bfv$ whose mean
vanishes. Averaging (\ref{1}) gives 
\beq
\partial_t \oS= -{\overline{ \bfv\cdot\nabla \cc }}.
\label{2}
\ee
To proceed, we need 
to derive an equation that expresses the right side of (\ref{2}) 
in terms of $\oS$. This requires the evolution equation for $\cc$. 
Subtracting (\ref{2}) from (\ref{1}) we obtain
\beq
\partial_t \cc = -\bfv\cdot \nabla \oS -\bfv \cdot \nabla \cc +{\overline{\bfv\cdot
\nabla \cc}}.
\label{3}
\ee
We are now faced with the need for a closure: 
If we  re-write the term on the right of  (\ref{2})
as $\int {\overline{ \bfv\cdot\nabla  \partial_t' \cc(t')}} dt'$,
then using (\ref{3}), a triple correlation 
of the form $\int {\overline { \bfv(t)\cdot\nabla (\bfv(t')\cdot \nabla \cc(t')) }}dt'$
arises.  In this approach, triple correlations are sometimes ignored 
by one of the following arguments: 
(i) $c<<\oS$, 
(ii) Gaussian statistics [e.g. see \cite{leslie,lesieur}] for which 
correlations of odd number of fluctuating quantities vanish, or  
(iii) the $\delta$-function-correlated in time 
[\cite{zeldovich88,zeldovich90,kraichnan68}] (or 
finite but short correlation time [\cite{elperin2001,fbc}])
%elperin2000
approximation.  These short correlation time approximations relate to 
the fact that upon one more iteration of the triple correlation by 
substituting $\cc(t')=\int \partial_{t''} \cc(t'') dt''$, and using (\ref{3}) again,
the triple correlations can be shown 
to contribute a term of order $\tau_c/\tau_{ed}$ 
times that of the double correlation  that arises from 
plugging the second term of (\ref{3}) back into 
(\ref{2}), where $\tau_c$ is the correlation time and 
$\tau_{ed}=1/v_2k_2$ is an eddy turnover time
associated with speed $v_2 = {\overline{\bfv^2}}^{1/2}/\sqrt 3$, 
and wavenumber, $k_2$, of the dominant turbulent scale.
If $\tau_c/\tau_{ed} < 1$  then triple correlations 
for quantities at two different times might be 
argued to be small.
Using one of the above three approximations, 
and after assuming isotropy, 
(\ref{2}) would become  a diffusion equation  
\beq
\partial_t \oS=(1/3)\int {\overline {\bfv(t)\cdot  \bfv(t')}}dt'\nabla^2\oS
\simeq (\tau_c/3){\overline{ \bfv^2 }}\nabla^2\oS \equiv\beta\nabla^2\oS,
\label{old}
\ee
where the similarity follows from replacing 
the time integral with multiplication by $\tau_c$ and $\beta$ 
is the diffusion coefficient.
The solution for $\oS$ is then
\beq
\oS=\oS_0 {\rm Exp}[-(t /\tau_{ed})(\tau_c/\tau_{ed})(k_1^2/k_2^2)],
\label{soln1}
\ee
where $k_1$ is the wavenumber associated with $\oS$.  Eq. (\ref{soln1}) reduces to $\oS_0 {\rm Exp}[-(t /3\tau_{ed})(k_1^2/k_2^2)]$
as $\tau_c\rightarrow \tau_{ed}$.  Because we 
focus  on  Eulerian transport, 
obtaining (\ref{old}) required the significant approximations discussed 
above. (For Lagrangian transport, (\ref{old}) would 
follow more simply [e.g. 1-2,6-8].)
But each of the arguments that led to neglecting
Eulerian triple correlations is poorly justified. 
First, $c << \oS$ cannot be guaranteed 
at all times. Second, although Gaussian statistics of fluctuating quantities
enforce triple correlations to vanish exactly for any correlation time,
finite triple correlations are required in order for there
to be any nonlinear turbulent cascade [see \cite{leslie,mccomb}].
Third, $\tau_c/\tau_{ed} < 1$ is not guaranteed,
thus invalidating a universally justified expansion in this quantity.

To develop an improved 
Eulerian approach which avoids the above
weaknesses, we re-start with (\ref{2})
and instead take the { time-derivative} of $\partial_t\oS$ to obtain
\beq
\partial_t^2 \oS= -\partial_t{\overline  {\bfv\cdot\nabla \cc }} 
=-{\overline{ \partial_t \bfv\cdot\nabla \cc }}- {\overline { \bfv\cdot\nabla \partial_t \cc }}.
\label{5}
\ee
Using (\ref{3}) in the last term of (\ref{5}) we have
\beq
- {\overline {\bfv\cdot\nabla \partial_t \cc }}=
{\overline { v_i v_j }} \partial_i  \partial_j \oS 
+{\overline {v_i \partial_i v_j }}\partial_j \oS 
+{\overline{\bfv\cdot\nabla(\bfv\cdot\nabla)\cc}},
\ee
where the last term in (\ref{3})  does not contribute when placed
in the average.
The second term on the right vanishes for homogeneous incompressible turbulence.
The first term on the right, after isotropizing, gives the
diffusion term, and the last term is a triple correlation which
we denote by $T_c$.
Plugging the result back into (\ref{5}) we then have
\beq
\partial_t^2 \oS= -\partial_t{\overline {\bfv\cdot\nabla \cc }} 
=-{\overline{ \partial_t \bfv\cdot\nabla \cc }} + T_c
+{1\over 3}{\overline{\bfv^2 }}\nabla^2\oS.
\label{6}
\ee
We must now use the equation of motion
\beq
\partial_t{\bfv}=-\bfv\cdot \nabla \bfv -\nabla p, 
\ee
(where $p$ is the pressure, and we assume incompressibility 
and set density $\rho= 1$)
for the first term on the right of (\ref{6}).
Noting that $\nabla^2 p = -(\partial_i v_j)(\partial_j v_i)$
for incompressible flows, we see that 
$p$ and thus $\nabla p$ depend on quadratic functions of $\bf v$ and
its gradients.
Thus when $\nabla p$ is combined with the  $\bfv\cdot\nabla \bfv$ term 
into
$\partial_t\bfv$, the third term of (\ref{6})
contributes a triple correlation  $T_v$,
similar to the triple correlation $T_c$ in that it is 
composed of  2 powers of $\bfv$ and one power of $s$.
We combine the two triple correlations and write $T= T_c + T_v $. 
Defining $\oQ=-{\overline{\bfv\cdot\nabla \cc}}$,  
the coupled equations to be solved are then
\beq
\partial_t{\oS}=\oQ
\label{7}
\ee
and
\beq
\partial_t{\oQ}=
{1\over 3}{\overline{\bfv^2}} \nabla^2 \oS 
+ T={1\over 3}{\overline{\bfv^2}}\nabla^2 \oS -\oQ /\tau_d, 
\label{8}
\ee
where in (\ref{8}) we have replaced $T$ by a damping term
of the form $-\oQ/\tau_d$, 
where $\tau_d = f\tau_{ed}$ and $f$ is a dimensionless constant expected
to be $\sim 1$.  This replacement is our closure.
Note that since  $\oQ=-{\overline{\bfv\cdot\nabla \cc}} 
=-{\overline{\nabla\cdot (\bfv \cc)}}$,
for incompressible flows, Gauss' theorem then shows that 
$\oQ$ represents a negative flux of $s$ through the volume of the averaging.
%volume $\sim k_1^{-3}$.

This ``minimal $\tau$-approximation'' for $T$  and the value
of $f$ can be tested based on the implications we now derive. 
Combining (\ref{7}) and (\ref{8}),  we have 
\beq
\partial_t^2{\oS}+\partial_t\oS/\tau_d -{1\over 3}{\overline{\bfv^2}} \nabla^2 \oS =0.
\label{10}
\ee
If we assume solutions of the form $\oS=\oS_0{\rm Exp}[nt+i{\bf k}\cdot {\bf x}]$
then (\ref{10}) becomes
$
n^2+n/\tau_d+1/\tau_L^2=0,
$
where $1/\tau_L= k_1 {\overline{ v^2}}^{1/2}/\sqrt 3\simeq k_1 v_2$.
The solution is
%\beq
$n={1\over 2\tau_d}\left(-1 \pm \sqrt{1-4{\tau_d^2\over \tau_L^2}}\right)$.
%\ee

In the limit that $\tau_L < \tau_d/2$, which corresponds
to the case when triple correlations are assumed small, 
the solution becomes oscillatory. This  
would be unphysical if concentrations of passive
scalars do not oscillate. 
To order of magnitude, this regime can also be written
$k_1 > k_2/2$, which includes a regime
in which the turbulent scale can be larger than the mean
field scale, threatening the scale separation
of the mean field formalism.

We now consider the case $\tau_L \ge 2\tau_d$, 
or equivalently $k_1 \le k_2/2$.
If $k_1=k_2/2$, then $n\sim -1/(2\tau_d)$ as expected when
$k_1$ and $k_2$ are not widely separated.
The more physical limit is $k_1 << k_2$
so that  $\tau_L >> \tau_d$. In this case, we 
Taylor expand in $\tau_d^2/\tau_L^2$
to obtain
$n={1\over 2 \tau_d}\left(-1 \pm \left({1-{2\tau_d^2\over \tau_L^2}}\right)\right)$.
The ``+'' solution gives $n\simeq-\tau_d/\tau_L^2\sim - 
(f /\tau_{ed})( k_1^2/k_2^2)=(f^2 /\tau_{d})( k_1^2/k_2^2)$
which gives the expected time scale of diffusion for
a quantity of scale $1/k_1$, subject to turbulent motions
of scale $1/k_2$ when $f\sim 1$. The ``-'' solution, which is new, gives 
$n\simeq -1/\tau_d$. The total
solution can be written
\beq
\oS= \oS_{0,+}{\rm Exp}[- t(f^2/\tau_d) (k_1/k_2)^2]+\oS_{0,-}{\rm Exp}[- t/\tau_d].
\label{soln}
\ee
For $f\sim 1$, 
the first term on the right depends on the longer decay constant and 
will dominate for all finite $S_{0,+}$ at late times.
The second term on the right has a 
much shorter decay constant since $k_1 << k_2$ and
requires further analysis.
Toward this end, we solve for $\oS_{0,-}$ and $\oS_{0,+}$ 
for several choices of  initial  conditions for 
$\partial_t \oS|_{t=0}=\oQ(0)$ and  $\oS_0=\oS(t=0)=\oS_{0,+}+\oS_{0,-}> 0$.

First consider the case $\oS_0>0$ and $\partial_t \oS|_{t=0}=\oQ(0)=0$.
From (\ref{soln}) we then obtain $\oS_{0,+}=-(k_2/k_1)^2 \oS_{0,-}$
and thus $\oS_{0,+}= \oS_0/(1- k_1^2/k_2^2)$.
Then
\beq
\oS= {\oS_{0}\over(1-k_1^2/k_2^2)}{\rm Exp}[- t(f^2/\tau_d) (k_1/k_2)^2]-
{(k_1^2/k_2^2)\oS_{0}\over(1-k_1^2/k_2^2) }{\rm Exp}[- (t/\tau_d)].
\label{soln2}
\ee
The first term on the right dominates for all time,
and the solution is thus similar to the familiar diffusion solution
of (\ref{old}).
From (\ref{7}), the 
time derivative of this solution gives  
%(\ref{soln}) dominates for all time.  We have 
%\beq
%\oS= \oS_{0}{\rm Exp}[- (t/\tau_d) (k_1/k_2)^2],
%\label{soln2}
%\ee
\beq
\partial_t\oS= \oQ=
-{f^2(k_1^2/k_2^2)\oS_{0}\over \tau_d(1-k_1^2/k_2^2)}{\rm Exp}[- t(f^2/\tau_d) (k_1/k_2)^2]+
{(k_1^2/k_2^2)
\oS_{0}\over\tau_d(1-k_1^2/k_2^2) }{\rm Exp}[- (t/\tau_d)].
\label{soln2d}
\ee
At early times, the flux 
$\oQ$ will begin to deviate from zero. In particular, 
the second term on the right decays more rapidly than the first
so $\oQ$ evolves to be negative.
This corresponds to a flux of $s$ from outside to inside the volume over
which the averaging is performed.

Let us now consider the case in which  $\oQ$ is initially finite, meaning that
there is an initial  net flux of $c$ in or out of the averaging volume.
More specifically, we consider $\oS_0=0$ and  $\partial_t \oS|_{t=0}=\oQ(0)\ne 0$.
We then obtain $\oS_{0,-}= -\oS_{0,+}=-\tau_d\oQ(0)/(1-k_1^2/k_2^2)$.
Then
\beq
\oS= {\tau_d \oQ(0)\over (1-k_1^2/k_2^2)}{\rm Exp}[- t(f^2/\tau_d) 
(k_1/k_2)^2]-{\tau_d\oQ(0)\over (1-k_1^2/k_2^2)}{\rm Exp}[- t/\tau_d],
\label{soln3d}
\ee
again we see that  $\oS$ diffuses in the usual way;
the first term on the right dominates for all time just as in the
previous case. 
If $\oQ(0)>0$ ($\oQ(0)<0$), then $\oS$ becomes increasingly less positive 
(negative) with time, asymptoting to zero.  Taking the time derivative 
gives 
\beq
\oQ= -{f^2 (k_1/k_2)^2\oQ(0)\over (1-k_1^2/k_2^2)}{\rm Exp}[- t(f^2/\tau_d) 
(k_1/k_2)^2]+{\oQ(0)\over (1-k_1^2/k_2^2)}{\rm Exp}[- t/\tau_d].
\label{soln3e}
\ee
For 
$t < -2\tau_d {\rm Ln}[(f (k_1/k_2)]/[1- f^2(k_1^2/k_2^2)]$ 
and $k_1/k_2 << 1$,  
(\ref{soln3e}) becomes 
%\beq
$
\oQ\simeq \oQ(0){\rm Exp}[- t/\tau_d].
$
%\label{soln3a}
%\ee
For $\oQ(0)> 0$  ($\oQ(0)< 0$) 
the initial net flux 
of $s$ is outward (inward) and the
solution thus implies that this outward  (inward) flux would decay on a
time scale $\sim \tau_d$ before the first term on the right
of (\ref{soln3e}) takes over. Then $\oQ$ changes
sign, becomes negative (positive) and eventually diffuses on
the same time scale that $\oS$ diffuses.

For $f\sim 1$, it is noteworthy that in the above two sets of solutions
for different initial conditions, $\oS$ essentially
diffuses as it would if it it were governed by (\ref{old}), the
``textbook'' equation for mean scalar diffusion.
But the evolution of $\oQ$ in our approach reveals more subtle properties
of the transport process. For an initially finite $\oQ$ and 
$\oS_0=0$,  
the fact that $\oQ$ incurs an initial  decay in the presence of isotropic 
turbulence on a time scale $\sim\tau_{ed}$ 
can be understood physically: Without an initial  mean field such as $\oS$, 
no net flux of $s$ can be expected to survive longer than an eddy 
turnover time for isotropic nonhelical turbulence since 
the assumption of statistical isotropy would be otherwise violated.  
%Indeed we found that for $\oS_0=0$ and $\oQ\ne 0$, 
%an initially positive (negative) flux thus decays rapidly. 
But as $\oS$ becomes positive (negative), 
$\oQ$ evolves to represent an inward (outward) flux of $s$ 
associated with the diffusion of $\oS$ and then evolves on the same 
time scale as $\oS$.

Further insight into the closure $T=\oQ/\tau_d$
comes from thinking qualitatively about entropy. 
A state with $\oS=0$ but large magnitude of $\oQ$  
can  be thought of as one with lower entropy than one in which $\oQ$ has a 
small magnitude, since a large $\oQ$ represents a 
macroscopically ordered anisotropic state.  Increasing the entropy with time 
would mean decaying $\oQ$.
This motivates a closure in which 
$T$ is replaced by a term representing decay of $\oQ$ rather than growth. 
The precise relation between $T$ and $\oQ$ can be tested numerically.

Our closure is akin to the ``$\tau$ approximation''
used by [\cite{kleeorin96}] for the anisotropic part 
of turbulent quantities. 
We call our closure the ``minimal $\tau$ approximation'' as
we use it for the full turbulent quantities.
This closure is simpler than 
the Eddy Damped Quasi Normal Markovian (EDQNM) 
[\cite{lesieur,orszag70,pfl}] and the Markovian Random Coupling Model (MRCM)
[\cite{frisch74}] closures. In these closures, 
$\partial_t T \propto \oQ^2$ rather than $T\propto \oQ$.
In the MRCM however, like ours, the decorrelation of triple moments
is taken to be time independent. For purposes of calculating the
diffusion of a mean scalar, we do not worry about the 
effect on the energy spectral index resulting from this choice.
Further numerical work should compare the relative
effectiveness of these various approaches.

Three other features of our approach are important to emphasize. 
First,  we did  not need to introduce a correlation time as  
correlations between quantities at different times never enter the theory. 
Second, the oscillations in the $\oS$ found when 
triple correlations were assumed to vanish highlight 
an likely unphysical consequence of any mean scalar transport 
theory which ignores terms providing the turbulent cascade. 
The latter is provided by finite 
triple correlations, which then also suppress these oscillations.
Third, with respect to the time evolution of $\oQ$, the
present approach is more useful than that implied by 
(\ref{soln1}). There, for $\oS_0=0$,  nothing happens 
without an additional source term, even if $\partial_t\oS|_{t=0}=\oQ(0)\ne 0$.
Thus, unlike the system of (\ref{7}) and (\ref{8}), (\ref{old}) 
does not model the evolution of a system in which a concentration of 
material is fed into a turbulent flow initially
devoid of material.

Our approach herein has now been 
shown to be consistent with fully 3-D numerical simulations
of turbulent diffusion of a mean scalar (\cite{b2003}). 
But the approach 
is also applicable to any equation for a mean quantity that depends
on the correlation of two fluctuating quantities.
A generic summary of the procedure is to replace the first order
equation of the quantity whose time derivative is of interest
by a second order equation by taking the derivative.
The derivative of the  double
correlation is then to be re-expressed using the known dynamical equations
for the time derivatives of each of the factors in the correlation,
and then one takes the triple correlations that arise to be linearly 
proportional to double correlations. 
The coupled system of equations must then be solved.
The approach has also been  successful [\cite{bf02}]
in mean field electrodynamics, where the time evolution for the 
mean magnetic helicity 
depends on the curl of the turbulent electromotive force 
$\emfb={\overline{\bfv\ts\bfb}}$, where $\bfb$ is the fluctuating magnetic field
in Alfv\'en units.
%$\partial_t \bbB=\curl {\overline {\bfv\ts\bfb }} +\curl {\bbV\ts \bbB} +\lambda \nabl%a^2\bbB$.
The procedure involves taking the time-derivative
of $\emfb$, using the equation of motion for $\partial_t \bfv$ and 
the induction equation  for  $\partial_t\bfb$, and replacing the 
resulting triple correlations by a damping term $\propto \emfb$.
The result, using $\tau_d\sim \tau_{ed}$ ($f\sim 1$)
is consistent [\cite{bf02}] with MHD simulations 
[\cite{b2001}].

\ni Thanks to A. Brandenburg for comments. EB acknowledges DOE grant DE-FG02-00ER54600.

\singlespace
%\begin{*}{thebibliography}{}
\enumerate

\bibitem{taylor21} G.I. Taylor, ``Diffusion by Continuous Movements,''
Proc. Lond. Math. Soc., {\bf 20} 196 (1921)

\bibitem{monin} A.S. Monin \& A.M. Yaglom, {\it Statistical Fluid
Mechanics} (MIT, Cambridge MA, 1975).

\bibitem{zeldovich88} Y.B. Zeldovich, A.A. Ruzmaikin, 
S.A. Molchanov, D.D. \& Sokoloff, 
``Self-Excitation of a Nonlinear Scalar Field in a Random Medium,''
PNAS, {\bf 84}, 6323 (1987)

%\bibitem{zeldovich88} Y.B. Zeldovich, A.A. Ruzmaikin, 
%S.A. Molchanov, D.D. \& Sokoloff, Sov. Sci. Rev. C. Math. Phys., 
%{\bf 7}, 1 (1988)

\bibitem{zeldovich90} Y.B. Zeldovich, A.A. Ruzmaikin, 
D.D. \& Sokoloff, {\it The Almighty Chance}, (World Scientific, Singapore 1990)

\bibitem{leslie}    D.C. Leslie  {\it
Developments in the Theory of  Turbulence}, (Clarendon, Oxford, 1983) 

\bibitem{mccomb}    W.D. McComb,  {\it
Physics of Fluid Turbulence}, (Clarendon, Oxford, 1990) 

\bibitem{lesieur}    M. Lesieur,  {\it
Turbulence in Fluids}, (Kluwer Press, Dordrecht, 1997) 

\bibitem{frisch} 
   U. Frisch,  {\it Turbulence}, (Cambridge University Press, Cambridge, 1995) 

\bibitem{shraiman} B.L. Shraiman \& E.D. Siggia, {\it Scalar Turbulence}
Nature {\bf 405}, 639 (2000)

\bibitem{falkovich} G. Falkovich, G., K. Gaw{\c e}dzki, \& 
M. Vergassola, ``Particles and fields in fluid turbulence,'' 
Rev. Mod. Phys., {\bf 73}, 913 (2001)

%\bibitem{elperin2000} T. Elperin, N. Kleeorin,  
%I. Rogachevskii,  \& D. Sokoloff,  PRE, {\bf 61}, 2617(2001)

\bibitem{elperin2001} 
T. Elperin, N. Kleeorin,  
I. Rogachevskii,  \& D. Sokoloff,  
``Mean-field theory for a passive scalar advected by a turbulent velocity field with a random renewal time,'' PRE, {\bf 64}, 26304 (2001)

\bibitem{avellaneda92} M. Avellaneda \& A.J. Madja,
Commun. Math. Phys., 
``Mathematical Models with Exact Renormalization for Turbulent
Transport,'' {\bf 146}, 381 (1992)

%\bibitem{avellaneda94} M. Avellaneda \& A.J. Madja,
%Philos. Trans. R. Soc. Lond., Ser. A., {\bf 346}, 205 (1994)

\bibitem{pasquill83}F. Pasquill \& F.B. Smith,
{\it Atmospheric Diffusion},  (Ellis Horwood, Chichester, 1983).

\bibitem{hodgson98} L.S. Hodgson, 
\& A. Brandenburg, 
``Turbulence effects in planetesimal formation,''
Astron. \& Astrophys., {\bf 330}, 1169 (1998)

\bibitem{baldyaga99}J. Baldyaga \& J.R. Borne,
{\it Turbulent Mixing and Chemical Reactions}
(Wiley, Chichester, 1999).

\bibitem{bf02} E.G. Blackman \& 
G.B. Field, ``New Nonlinear Mean Field Dynamo and Closure Theory,''
PRL, {\bf 89}, 265007 (2002)

\bibitem{kraichnan68} R.H. Kraichnan,
''Convergents to Infinite Series in Turbulence Theory,''
Phys. Rev., {\bf 174} 240 (1968)

\bibitem{fbc}
G.B. Field, E.G. Blackman \& H. Chou, ApJ, 
``Nonlinear Alpha Effect in Dynamo Theory,''
{\bf 513}, 638 (1999)

\bibitem{kleeorin96} 
N. Kleeorin, M. Mond, I. Rogachevskii, 1996, A. \& Ap.,  
``Magnetohydrodynamic turbulence in the solar convective zone as a source of oscillations and sunspots formation''
{\bf 307} 293 (1996) 

\bibitem{orszag70} S. Orszag, J. Fluid Mech., 
``Analytic Theories of Turbulence,'' {\bf 41}, 363 (1970)

\bibitem{pfl} 
 A. Pouquet, U. Frisch, \& J. Leorat,  
``Strong MHD helical turbulence and the nonlinear dynamo effect,''
JFM, {\bf 77}, 321 (1976)

\bibitem{frisch74} U. Frisch, M. Leiseiur, A. Brissaud,
``Markovian Random Coupling Model for Turbulence''
JFM 65 {\bf 145} (1974)

%\bibitem{pouquet78} A. Pouquet JFM {\bf 88} 1 (1978)

%\bibitem{das91} A.K. Das, J. Appl. Phys. {\bf 70} 1355 (1991).

\bibitem{b2001}
A. Brandenburg, 
``The Inverse Cascade and Nonlinear Alpha-Effect in Simulations of Isotropic Helical Hydromagnetic Turbulence,'' ApJ, {\bf 550}, 824 (2001)

\bibitem{b2003}
A. Brandenburg, 
``Non-Fickian Diffusion and the tau-approximation from
numerical turbulence,'' astro-ph/0306521,
submitted to Phys. Rev. E. 2003.

%\bibitem{krause}   F. Krause \& K.-H. R\"adler, 
%{\it Mean-field Magnetohydrodynamics and Dynamo Theory}, 
%(Pergamon Press, New York, 1980)

%\bibitem{parker79}  
%E.N. Parker, {\it Cosmical Magnetic Fields}, (Oxford: Clarendon
%Press, 1979)

%\end{thebibliography}
\bigskip
\bigskip
\bigskip
\bigskip
\end{document}